# Basic concepts in single-molecule electronics

C. J. Lambert

Dept. of Physics, Lancaster University, Lancaster, LA1 4YB, U.K.

This tutorial outlines the basic theoretical concepts and tools which underpin the fundamentals of phase-coherent electron transport through single molecules. The key quantity of interest is the transmission coefficient T(E), which yields the electrical conductance, current-voltage relations, the thermopower S and the thermoelectric figure of merit ZT of single-molecule devices. Since T(E) is strongly affected by quantum interference (QI), three manifestations of QI in single-molecules are discussed, namely Mach-Zehnder interferometry, Breit-Wigner resonances and Fano resonances. A simple MATLAB code is provided, which allows the novice reader to explore QI in multi-branched structures described by a tight-binding (Hückel) Hamiltonian. More generally, the strengths and limitations of materials-specific transport modelling based on density functional theory are discussed.

## A. Introduction

The aim of this brief tutorial is to provide an introduction to basic theoretical concepts and tools, which pervade the field of room-temperature single-molecule electronics. The technology goal of the field is to enable the transformative development of viable, molecular-based and molecularly-augmented electronic devices. Such devices would enable new sensor technologies for pressure, acceleration and radiation, along with chemical sensors capable of detection and analysis of a single molecule. They could also lead to new designs for logic gates and memory with orders of magnitude reduction of both power requirements and footprint area, and new thermoelectric devices with the ability to scavenge energy with unprecedented efficiency. However, realisation of these benefits of single-molecule electronics will require the unprecedented assembly of bespoke molecules in reproducible and scalable platforms with multiple electrodes separated by molecular length scales. At present, the required level of control is far beyond current technologies and research on single-molecule electronics is focussed mainly on the fundamental science that will underpin future developments.

One example of such fundamental science is the room-temperature observation of quantum interference (QI) in single molecules connected to nano-gap electrodes. At a fundamental level, almost all anticipated performance enhancements associated with single-molecule electronics are derived from transport resonances associated with QI. Quantum interference controls and is controlled by molecular conformation, charge distribution and the energies of frontier orbitals. The symbiotic relationship between quantum interference and physical and electronic structure, which are sensitive to environmental factors and can be manipulated through conformational control, polarisation or redox processes (electrical vs. electrochemical gating) leads to new opportunities for controlling the electrical and thermoelectrical properties of single molecules connected to nano-electrodes. Room-temperature QI in single molecules was demonstrated experimentally only recently[1-9] and it is of interest to identify new strategies for exploiting QI in technologically-relevant platforms.

The tutorial below is intended to provide the background concepts and theoretical tools, which will inform the development of such strategies. The tutorial is based on notes and examples given to early-stage researchers in my group and assumes only the most basic training in mathematics.

## B. The standard model of phase-coherent electron transport through single-molecules.

To begin, this tutorial I shall briefly outline a 'standard model' of single-molecule electronics, both from a conceptual and computational viewpoint. The glossary below summarises some of the key concepts and ingredients pervading current research in room-temperature single-molecule electronics, including three types of resonances arising from QI, namely Breit-Wigner, Fano and Mach-Zehnder resonances.

## Glossary

*Molecular junction* A molecular junction is a molecular electronic device in which a (single) molecule is chemically anchored to two conducting electrodes.

*Conjugated:* A system of alternating single and double bonds, which promotes extended electronic states, e.g. molecule **1** (right).

*Cross-conjugated:* An example of this is molecule **2** (right) in which π-system A is conjugated to B, and C to B, but A and C are not conjugated to each other.

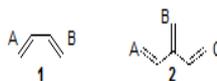

A *Breit-Wigner resonance* occurs when the energy $E$ of an electron passing through a molecule resonates with a backbone state of the molecule

A *Fano resonance* occurs when $E$ coincides with the energy of a bound state located on a pendant group, which couples to a backbone state

A *Mach-Zehnder resonance* occurs when partial de Broglie waves of energy $E$ traversing different paths interfere constructively, eg. in a multi-branch molecule.

*Anchor:* Traditionally these have been single point contacts, such as a thiol (S-H) or amine (NH$_2$). Spatially-extended anchors will bind via non-covalent π-π stacking and van der Waals interactions to electrode surfaces, with binding energies increasing with the area of overlap. Large planar anchors are more resilient to atomic-scale fluctuations in the electrode surfaces. Anthracenes, pyrenes, porphyrins and endohedral fullerenes are candidate anchors.

*Linker:* Ideally rigid, length-persistent and conjugated (ie possessing delocalised frontier orbitals) and with a small HOMO-LUMO gap, so the conductance decays only slowly with length. Porphyrin, polyyne and oligo(phenylene ethynlene) based structures are examples.

*Functional unit:* An electronically or otherwise active molecular building block that lends additional function to a molecular component

*Simple wire:* Two anchors connected by a rigid, length-persistent linker, potentially exhibiting Breit-Wigner resonances.

*A central functional unit* connected to two anchors by linkers, for example potentially exhibiting switchable Breit-Wigner resonances.

*A pendant unit* connected to a central functional unit. Can be designed to give Fano resonances and enhanced thermoelectrical properties, e.g. cross-conjugated molecules, or to be electrically inert so it can act as an insulator to a nearby gate.

Multiple-paths lead to Mach-Zehnder resonances, which can be controlled by preferentially gating or otherwise switching just one of the branches

Central functional unit can be shielded from the environment by an *in-built inert organic shell*, thereby protecting phase-coherence or shorting to a gate electrode.

Analyte molecules attracted to functional units can modify interference leading to few-molecule sensing. This effect is amplified by using a gate to move molecular resonances close to the Fermi energy. Alternatively a pre-existing solvation shell can be modified by a change in environment, leading to new sensing strategies.

Not shown in the glossary are the electrodes, which make electrical contact to the anchor groups on the left and right ends of the molecules. Usually these are made from gold and contact is made using a scanning –probe microscope of a mechanically-controlled break junction. However gold is rather mobile and forbidden from a CMOS lab. Consequently other electrode materials have been explored, including Pt, Pd, carbon nanotubes and graphene[9]. Indeed recently, Si-based platforms for molecular electronics have begun to be explored[10].

To illustrate the standard approach to computing the electrical conductance of single molecules, consider the molecule shown below, located between two gold electrodes.

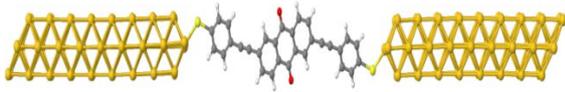

**Figure 1.** A molecule in contact with two gold electrodes.

The electrodes are connected to external reservoirs (not shown), which feed electrons of energy E into the gold electrodes. The energy distribution of electrons entering the left (right) gold lead from the left (right) reservoir is $f_{left}(E)$ ($f_{right}(E)$) and according to the Landauer formula, the current passing from left to right is

$$I = \left(\frac{2e}{h}\right)\int_{-\infty}^{\infty} dE\, T(E)[f_{left}(E) - f_{right}(E)] \quad (1)$$

where *T(E)* is the transmission coefficient for electrons passing from one gold lead to the other via the molecule. Clearly I=0 when $f_{left}(E) = f_{right}(E)$, because only differences in the distributions contribute to the net current.

Close to equilibrium,

$f_{left}(E) = [e^{\beta(E-E_F^{left})} + 1]^{-1}$ and $f_{right}(E) = [e^{\beta(E-E_F^{right})} + 1]^{-1}$, where $E_F^{left}$ ($E_F^{right}$) is the Fermi energy of the left (right) reservoir and $\beta = \frac{1}{k_B T}$, where T is the temperature. If V is the voltage difference between the left and right reservoirs, then $E_F^{left} = E_F + \frac{eV}{2}$ and $E_F^{right} = E_F - \frac{eV}{2}$. This means that at zero temperature, but finite voltage

$$I = \left(\frac{2e}{h}\right)\int_{E_F-eV/2}^{E_F+eV/2} dE\, T(E) \quad (2)$$

Consequently, the conductance *G=I/V* is obtained by averaging T(E) over an energy window of width eV centred on the Fermi energy. On the other hand, if *T(E)* does not vary significantly over an energy range of order eV, then the Fermi functions can be Taylor expanded to yield the electrical conductance in the zero-voltage, finite temperature limit:

$$G = I/V = G_0 \int_{-\infty}^{\infty} dE\, T(E)\left(-\frac{df(E)}{dE}\right) \quad (3)$$

where $G_0$ is the quantum of conductance,

$$G_0 = \left(\frac{2e^2}{h}\right)$$

Since the quantity $-df(E)/dE$ is a normalised probability distribution of width approximately $k_B T$, the above integral represents a thermal average of the transmission function T(E) over an energy window of width $k_B T$.

Finally, in the limit of zero voltage and zero temperature, one obtains

$$G = G_0 T(E_F) \qquad (4)$$

The transmission coefficient $T(E)$ is a property of the whole system comprising the leads, the molecule and the contact between the leads and the molecule. Nevertheless, if the contact to the electrodes is weak, then a graph of $T(E)$ versus $E$ will reflect the energy-level structure of the isolated molecule. In particular if the isolated molecule has energy levels $E_n$, (where $n$ represents the quantum numbers labelling the energy levels), then $T(E)$ will possess a series of peaks (ie resonances) located at energies in the vicinity of the levels $E_n$. These resonances will be discussed in more detail below.

The above expressions assume that during the time it takes for an electron to pass from one lead to the other, the system of leads plus molecule can be described by a time-independent mean-field Hamiltonian $H$. This means that an electron remains phase coherent as it passes from one gold lead to the next and does not undergo inelastic scattering. For short molecules, there is evidence that this is a reasonable assumption[11]. However as the length of a molecule increases, the probability of inelastic scattering (eg from phonons or other electrons) becomes non-negligible[12] and the above expressions require modification. In what follows, we shall confine the discussion to systems where the above expressions are valid.

On the other hand, just because a Hamiltonian is static on the timescale required to pass through the molecule, does not mean that the system is completely stationary, because on a longer timescale, the molecule may change its shape, the electrodes may distort or the environment surrounding the molecule may change. In this case, in the spirit of the Born-Oppenheimer approximation, one should construct an ensemble of Hamiltonians describing each realisation of the system (ie. electrodes, molecule and environment) and compute $T(E)$ for each case. Since the environment and molecular conformation can fluctuate many times on the timescale of a typical electrical measurement, an ensemble average of such transmission functions is required. This ensemble average (ie average over snapshots) does not include the even slower changes in $H$ associated with the pulling apart of electrodes in a break-junction experiment, or with the changes associated with the switching on of a dc bias or gate voltage. In the presence of such slow changes, an ensemble average should be carried out at each electrode separation or gate voltage.

Starting from an arbitrary instantaneous mean-field Hamiltonian $H$ in a localised basis, the problem of computing the transmission coefficient $T(E)$ was solved in [13] and therefore the main problem is how to obtain $H$. Unfortunately there is no "text book of Hamiltonians" and theoreticians have to resort to a range of approximations and tricks to construct reasonable models. In the paper[13] and in other papers of that decade, the strategy used to obtain a Hamiltonian was to fit the matrix elements of a tight-binding Hamiltonian to band structures of the materials of interest. This is perfectly respectable, but it does not solve the problem of modelling interfaces between different materials. At that time, a commonly-used approach was to approximate the matrix elements describing an interface between two materials by some average of the matrix elements of the individual materials. Clearly there is no microscopic justification for such an approach. In the case of single-molecule electronics, an accurate description of interfaces between the electrodes and the molecule is vital and therefore an alternative approach was necessary. To overcome this problem, the SMEAGOL code[14,15] (Spin and Molecular Electronics in Atomically-Generated Orbital Landscapes) was created to compute $T(E)$ from the mean-field Hamiltonian of the density functional theory (DFT) code SIESTA and indeed to modify the SIESTA Hamiltonian in the presence of a finite bias. Similar strategies have been adopted by many groups using a range of DFT codes, including TRANSIESTA[16] and TURBOMOLE[17]. Recently a next-generation code (GOLLUM) has been released[18], which is freely distributed and additionally describes the environmental effects due to a surrounding solvent or gaseous atmosphere, the evolution of electrical properties during the pulling apart of electrodes in break junction experiments and a range of other more exotic effects associated with thermoelectricity, superconductivity and magnetism.

The above DFT-based approach is widely adopted and can be considered a 'standard model' of phase-coherent electron transport through single-molecules. The good news is that DFT can predict ground state energies, binding energies, molecular conformations, trends in HOMO-LUMO gaps in homologous series of molecules, symmetries of orbitals and a range of other properties (see below). The bad news is that DFT is not an exact theory and the underlying Hamiltonian cannot be relied upon to reproduce essential quantities such as the positions of the HOMO and LUMO levels relative to the Fermi energies of electrodes. There are many reasons for this lack of reliability, ranging from approximations associated with the underlying density functional, the fact that Kohn-Sham eigenvalues are not the energy levels of a system and the presence of mirror charges and screening due to the nearby electrodes[19].

To illustrate this deficiency of DFT, consider the theoretical predictions for the HOMO-LUMO gaps of simple linear carbon chains (oligoynes) with various end caps, shown in figure 2. In the limit that the number $2n$ of carbon atoms in the chain tends to infinity (ie in the limit $1/n$ tends to zero), these should all agree, but clearly this is not the case. A SIESTA LDA implementation of DFT yields a much-too-low value of less than 1eV. For comparison, figure 2 shows the result of a GW many-body calculation[20], which for an infinite polyyne chain yields a HOMO-LUMO gap of 2.2eV, which is close to the literature consensus of experimental values, shown in figure 3.

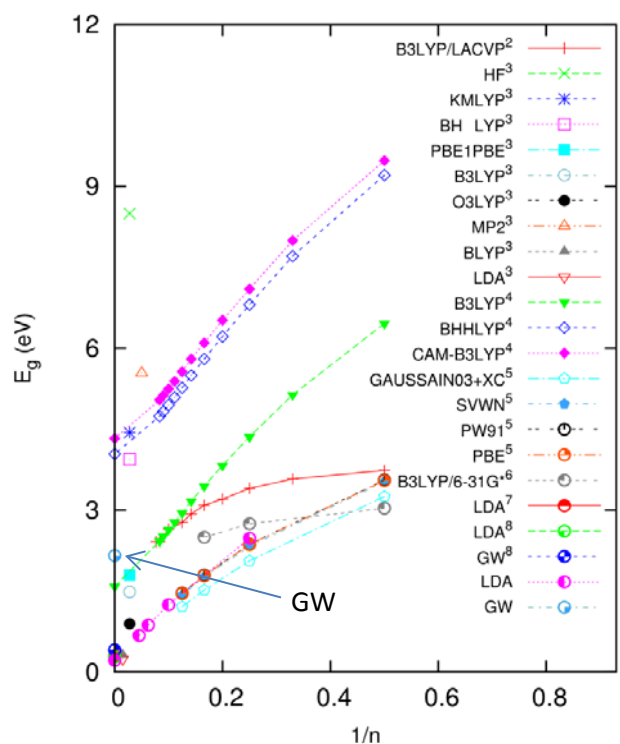

Figure 2 Theoretical values of the HOMO-LUMO gap (Eg) of oligoynes and of $1/n$ (where $n$ equals the number of carbon pairs) using various methods. The last two items of data correspond to recent results from my own group using LDA in the SIESTA code and GW in VASP. For references to original data, see the SI of [20].

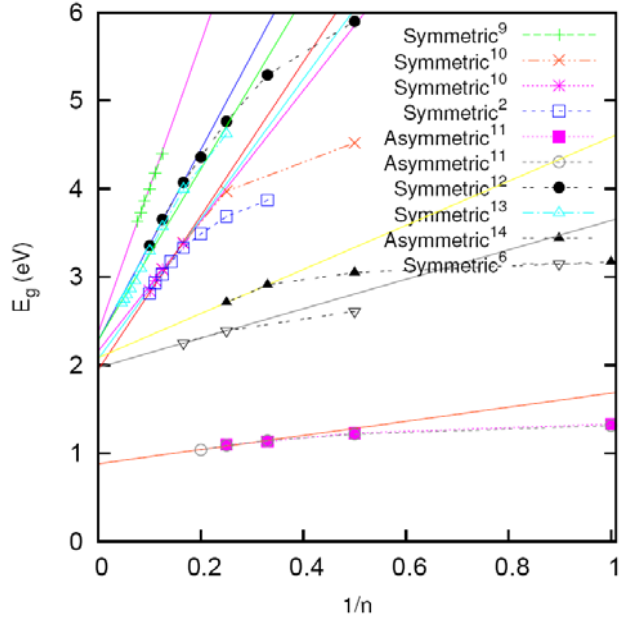

Figure 3. Experimental values of the HOMO-LUMO gap of short oligoynes as a function of $1/n$ (where $n$ equals the number of carbon pairs) for different molecules terminated with symmetric and asymmetric arrangements of end groups. Linear extrapolation to the limit of infinite oligoyne length estimates the band gap of polyyne between 2.0 and 2.3 eV in the majority of these measurements, which is in excellent agreement with our GW results. For references to original data, see the SI of [20].

Despite the superior performance of GW, DFT remains a widely-used workhorse, because with currently-available computers, GW can handle only small numbers of atoms and cannot describe many of the molecules of interest for single-molecule electronics.

To accommodate the inadequacies of DFT, a DFT mean-field Hamiltonian should be adjusted to accommodate known experimental facts or outputs from more accurate many-body simulations. In particular if the HOMO-LUMO (H-L) gap $E_G = E_L - E_H$ is known experimentally, then a scissor correction[21,22] should be implemented so that the corrected Hamiltonian reproduces the known value. In addition if the locations of the HOMO and LUMO levels $E_H$, $E_L$ relative to the Fermi energy $E_F$ of the electrodes are known, then the relative position of the Fermi energy should be adjusted to correct value. Once the two energy differences $E_L - E_F$ and $E_H - E_F$ are fixed by experiment, then the resulting phenomenological Hamiltonian and associated predictions for transport properties are probably the most accurate obtainable. This approach mirrors that used widely by the mesoscopics community to describe transport through quantum dots, where charging energies, electron affinities and ionisation potentials are usually fixed to agree with experiment and rarely computed from first principles. Useful information for locating the energy differences $E_L - E_F$ and $E_H - E_F$ includes the shape of current-voltage plots, the sign of the thermopower (see below), spectroscopic data and the temperature dependence of the electrical conductance. Trends in measured properties of homologous series of molecules and conductance changes due to electrostatic gating or the surrounding environment also provide valuable information for locating $E_F$. For example, systematic experimental and theoretical studies with thiol-terminated aliphatic and aromatic molecular wires confirm that electrons can remain phase coherent when traversing single molecules, with a non-resonant tunneling process as the main transport mechanism. The tunneling decay parameter β in $G = A \cdot e^{-\beta L}$, with G and L as the molecular conductance and length, respectively, can be as large as (0.4 Å$^{-1}$) for oligo-*para*-phenylenes[23] and (0.2 Å$^{-1}$) for oligophenyleneethynylenes[24,25]. Smaller β values are found for oligothiophenes (β = 0.1 Å$^{-1}$)[26], oligoynes (β = 0.06 Å$^{-1}$)[27], and Zn-porphyrin-containing wires[28]. Such values of β is can help to determine the position of the Fermi energy relative to the LUMO or HOMO levels. In principle, such account for the effect of image charges in the electrodes, which are known to renormalize $E_L$- $E_F$ and $E_H$ - $E_F$ by up to several hundred meV[19].

## C. Influence of quantum interference on electron transport through single-molecules.

Having dwelt upon the skeletons in the cupboard of DFT, I now discuss some concepts which underpin descriptions of QI within single molecules and which allow us to interpret the outcomes of complex material-specific DFT-based calculations. Since QI takes many forms, it is worth noting that QI in isolated molecules is distinct from QI in molecules connected to conducting electrodes. In the former, the system is closed and constructive QI leads to the formation of molecular orbitals $\{|\varphi_n\rangle\}$ and a discrete energy spectrum $\{E_n\}$, obtained by solving the Schrodinger equation $H|\varphi_n\rangle = E_n|\varphi_n\rangle$. In the latter, the system is open and the energies $E$ of electrons entering a molecule form a continuum. In this case, once the Hamiltonian is known, we are interested in calculating the transmission coefficient $T(E)$ of the molecule, which describes the transmission of de Broglie waves of energy $E$, which enter themolecule from the electrodes. The transmission coefficient $T(E)$ can exhibit both constructive and destructive interference, whereas the orbitals $\{|\varphi_n\rangle\}$ are formed from constructive interference only.

In an open system, QI should be described in terms of the interference pattern of de Broglie waves due to electrons traversing a molecule from one contact to another. The technical term for such an interference pattern is a Greens function. As a simple analogy, imagine a singer at the entrance to an auditorium, emitting a note of constant frequency. If the note is sustained for long enough this will set up a standing wave within the auditorium, with a complex pattern of nodes and antinodes. When the singer is located at position i, then the amplitude of the standing wave at position j is proportional to the Greens function G(j,i). If a window is opened at position j and the location of the window coincides with an antinode of the standing wave G(j,i), a loud sound will exit from the window. This is an example of constructive interference. On the other hand, if the window is located at a node, no sound will exit and destructive interference is said to occur. In this analogy, the auditorium and the sound wave are analogues of a molecule and a de Broglie wave respectively and in a model based on atomic orbitals, i and j label atomic orbitals at specific locations within a molecule. In this case, G(j,i) is the amplitude of a de Broglie wave at j, due to a source at i and is a property of the isolated molecule. If an electrode is attached to i=1 and a second electrode attached to j=2, then the transmission coefficient $T(E)$ will vanish when G(2,1)=0; ie when atomic orbital 2 coincides with a node.
In what follows, four examples of QI in molecules are described by analysing in detail a simple tight-binding model of $\pi$ orbitals. Since this is a tutorial introduction, many mathematical steps are included, so the reader should be able to reproduce all results with relative ease.

**Example 1. Mach-Zehnder interferometers**

As a first example of QI and to further contrast QI in open systems with QI in closed systems, figure 4a shows an example of a tight-binding ring with 6 atomic orbitals (sometimes referred to as 'atoms' or 'sites' for brevity), labelled *j*=1,2, … 6, described by a Schrodinger equation

$$\varepsilon_0\varphi_1^n - \gamma\varphi_6^n - \gamma\varphi_2^n = E_n\varphi_{j1}^n$$
$$\varepsilon_0\varphi_j^n - \gamma\varphi_{j-1}^n - \gamma\varphi_{j+1}^n = E_n\varphi_j^n \quad (j = 2,3,4,5) \quad (5)$$
$$\varepsilon_0\varphi_6^n - \gamma\varphi_5^n - \gamma\varphi_1^n = E_n\varphi_{j1}^n$$

In this simple tight-binding (Hückel) model, $\varphi_j^n$ is the amplitude of the nth molecular orbital on atom *j* and $E_n$ is the energy of the *n*th molecular orbital. The parameters $\varepsilon_0$ and $-\gamma$ are "site energies" and "hopping elements" determined by the type of atom in the rings and the coupling between the atoms. To solve this set of equations, it is illuminating to first consider the Schrodinger equation of an infinite chain of such atoms, which takes the form

$$\varepsilon_0 \varphi_j - \gamma \varphi_{j-1} - \gamma \varphi_{j+1} = E\varphi_j \quad (-\infty < j < \infty) \quad (6)$$

The solution to this equation is $\varphi_j = e^{ikj}$, where $-\pi < k < \pi$. Substituting this into equation 5b yields $(\varepsilon_0 - \gamma e^{-ik} - \gamma e^{ik})e^{ikj} = Ee^{ikj}$, so after cancelling $e^{ikj}$ on both sides, one obtains the dispersion relation

$$E = \varepsilon_0 - 2\gamma \cos k$$

This means that the 1-d chain possess a continuous band of energies between $E = \varepsilon_0 - 2\gamma$ and $E = \varepsilon_0 + 2\gamma$. More generally, the solution to equation 5b is $\varphi_j = e^{ikj} + \alpha e^{-ikj}$, where α is an arbitrary constant. For such an open system, the problem is not to compute E, since any value within the band is allowed. Instead we choose E and compute properties of interest as a function of E. For example, the dimensionless wave vector $k$ is given by

$$k(E) = cos^{-1}[(\varepsilon_0 - E)/2\gamma] \quad (7)$$

and the group velocity is $v(E)/\hbar a$ where a is the atomic spacing, $\hbar$ is Planck's constant and

$$v(E) = \frac{dE}{dk} = 2\gamma \sin k(E) \quad (8).$$

(The latter expression illustrates why it is convenient to use $-\gamma$ rather than $\gamma$ to denote hopping elements, because with this notation, if $\gamma$ is positive, then $v(E)$ has the same sign as $k(E)$.)

In contrast with the open system described by equation 6, when the solution $\varphi_j = e^{ikj}$ is substituted into equations 5, the only 6 linearly independent solutions which also satisfy the first and last equations are found to be $\varphi_j = e^{ik_n j}$, where $k_n = \frac{2\pi n}{6}$, $n = -2, -1, 0, 1, 2, 3$, with energies $E_n = \varepsilon_0 - 2\gamma \cos k_n$. Since energies corresponding to $k_n$ and $-k_n$ are degenerate, the complex exponentials $e^{ik_n j}$ and $e^{-ik_n j}$ can be added or subtracted to yield the following 6 real solutions: $\varphi_j^n = \sin k_n j$ (where n=1,2) and $\varphi_j^n = \cos k_n j$ (where n=0,1,2,3). Clearly the levels $E_1$ and $E_2$ are doubly degenerate and the levels $E_0$ and $E_3$ are non-degenerate and in contrast with the open system of equation 6, E is no longer continuous.

This example can be considered to be a simple model of $\pi$ electrons in a phenyl ring, with one electron per orbital, in which case the 6 electrons occupy the lowest three levels up to the HOMO $E_H = E_1$ and the LUMO is $E_L = E_2$. Clearly the mid-point of the HOMO-LUMO gap is $\bar{E} = (E_H + E_L)/2 = \varepsilon_0$ and the parameters $\varepsilon_0, \gamma$ could be approximated by choosing $E_1$ and $-E_2$ to coincide with the ionisation potential and electron affinity of the molecule.

Now consider attaching a 1-dimensional lead to atom $i=1$ and a second 1-dimensional lead to a different atom $j$, to create an open system with two electrodes. If $j = 4$ (as shown in figure 4b), then this is known as a para-coupled ring, whereas if $j = 3$ (or 5), as shown in figure 4c, it is a meta-coupled ring. For $j = 2$, the coupling is 'ortho'. In physics, the optical analogue of such a structure is known as a Mach-Zehnder interferometer.

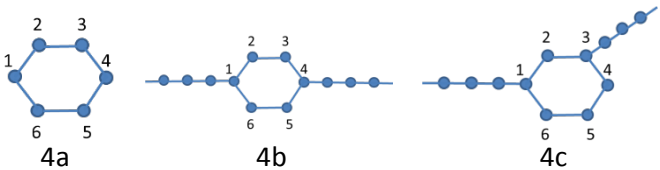

Figure 4. 4a shows a closed system, 4b shows an open system with para coupling to the leads and 4c shows an open system with meta coupling.

Since the leads are infinitely long and connected to macroscopic reservoirs (not shown), sytems 4b and 4c are open systems. In these cases, the transmission coefficient $T(E)$ for electrons of energy $E$ incident from the first lead is obtained by noting that the wave vector $k(E)$ of an electron of energy of energy E traversing the ring is given by $k(E) = cos^{-1}(\varepsilon_0 - E)/2\gamma$.

When E coincides with the mid-point of the HOMO-LUMO gap, ie when $E = \varepsilon_0$, this yields $k(E) = \pi/2$. Since T(E) is proportional to $|1 + e^{ikL}|^2$, where $L$ is the difference in path lengths between the upper and lower branches, one obtains constructive interference in the para case, (where $e^{ikL} = e^{i0} = 1$) and destructive interference in the meta case, (where $e^{ikL} = e^{i2k} = -1$).

More precisely, for the ring of N atoms, the Greens function of the isolated ring of N atoms (see example 4 below) is

$$G_{ring}(j,i) = A\cos k(|j-i| - N/2) \qquad (8)$$

where N=6 for the ring in figure 4a. In this expression, k is given by equation (7) and the amplitude *A* is

$$A = \frac{1}{2\gamma \sin k \sin kN/2} \qquad (9)$$

The transmission in the case of para coupling is determined by the Greens function $G_{ring}(4,1) = A$. In the case of meta coupling, it is determined by $G_{ring}(3,1) = A\cos k$. In the case of ortho coupling, it is determined by $G_{ring}(2,1) = A\cos 2k$.

In the centre of the HOMO-LUMO (H-L) gap, where k=π/2, these reduce to *A*, 0 and *–A* for the para, meta and ortho cases respectively, whichdemonstrates that para and orth correspond to constructive QI whereas meat corresponds to destructive QI.

As well as illustrating constructive and destructive QI, the above expressions also illustrate the non-classical manner in which conductances add in parallel. For example since figure 4b has two branches in parallel, the classical conductance of figure 4b should be twice that of a single branch. However for a linear chain of *N* atoms, the Greens function connecting atoms 1 and *N* at opposite ends of the chain is[29]

$$G_{chain}(M,1) = \frac{-\sin k}{\gamma \sin k(N+1)} \qquad (10)$$

which for *N*=2 reduces to 1/γ at the H-L gap centre. Hence at the gap centre,

$$\frac{G_{ring}(2,1)}{G_{chain}(N,1)} = \frac{1}{2}$$

which means that the transmission of the para-ring at the gap centre is $(1/2)^2$ = ¼ of the transmission of a single linear chain. This demonstrates that QI must be respected when combining conductances at a molecular level.

Finally, for completeness it is worth noting that the above Greens functions are those of isolated rings and chains, not connected to leads. As shown in example 4, when the leads are attached, the Greens functions become

$$G_{ring}(j,i) = \frac{\cos k(|j-i| - N/2)}{2\gamma \sin k \sin \frac{kN}{2} + \sigma_{ring}} \qquad (11)$$

and

$$G_{chain}(M,1) = \frac{-\sin k}{\gamma \sin k(M+1) + \sigma_{chain}} \qquad (12)$$

where $\sigma_{ring}$ and $\sigma_{chain}$ are self energies due to the attachment of the leads. (Both $\sigma_{ring}$ and $\sigma_{chain}$ vanish in the limit that the coupling to the leads vanishes.)

**Example 2. The Breit-Wigner formula**

The above example shows that QI between multiple paths can lead to resonances or anti-resonances in the transmission function *T(E)* controlling charge transport and therefore if these peaks or dips in *T(E)* are rather narrow, large values of the slope *dT(E)/dE* can be obtained. If this occurs near the electrode Fermi energy, then the electrical conductance of the molecule can be extremely sensitive to changes in the environment, including changes in the surrounding atmosphere, the temperature, a gate voltage or changes in molecular conformation or electrode spacing. If such sharp features in $T(E)$ can be controlled, then the system may behave as a sensitive detector, transistor or an efficient thermoelectric device.

The ability to tune molecular orbitals ( eg by varying the parameters $\varepsilon_0$ and $-\gamma$ in equations 5) is an example of how *constructive* interference can be exploited to control transport, since molecular orbitals are a direct result of constructive interference within the isolated molecule. For electrons of energy $E$ passing through a single molecular orbital $\varphi_j^n$, the simplest description of constructive interference is provided by the Breit-Wigner formula[30]

$$T(E) = 4\Gamma_1 \Gamma_2/[(E - \varepsilon_n)^2 + (\Gamma_1 + \Gamma_2)^2], \qquad (13)$$

where $T(E)$ is the transmission coefficient of the electrons, $\Gamma_1$ and $\Gamma_2$ describe the coupling of the molecular orbital to the electrodes (labeled 1 and 2) and $\varepsilon_n = E_n - \Sigma$ is the eigenenergy $E_n$ of the molecular orbital shifted slightly by an amount $\Sigma$ due to the coupling of the orbital to the electrodes. This formula shows that when the electron resonates with the molecular orbital (ie when $E = \varepsilon_n$), electron transmission is a maximum. The formula is valid when the energy E of the electron is close to an eigenenergy $E_n$ of the isolated molecule, and if the level spacing of the isolated molecule is larger than $(\Gamma_1 + \Gamma_2)$. In the case of a symmetric molecule attached symmetrically to identical leads, $\Gamma_1 = \Gamma_2$ and therefore on resonance, when $E = \varepsilon_n$, $T(E)=1$. On the other hand in the case of an asymmetric junction, where for example $\Gamma_1 \gg \Gamma_2$, the on-resonance values of $T(E)$ (when $E = \varepsilon_n$) is less than unity.

Although the resonance condition ($E = \varepsilon_n$) described by the Breit-Wigner formula is a consequence of constructive interference, it is worth noting that the formula also contains information about destructive interference. As an example, for the case of 1-dimensional leads, $\Gamma_1$ is proportional to $(\varphi_i^n)^2$ and $\Gamma_2$ is proportional to $(\varphi_j^n)^2$, where $\varphi_i^n$ and $\varphi_j^n$ are the amplitudes of the wavefunction (ie molecular orbital) evaluated at the contacts and if either $\varphi_i^n$ or $\varphi_j^n$ coincide with nodes then $\Gamma_1$ or $\Gamma_2$ will vanish. For this simple example, g(j,i)= $\varphi_i^n \varphi_j^n/(E - \varepsilon_i)$ and clearly vanishes when $\varphi_i^n$ or $\varphi_j^n$ coincide with nodes. More generally, $\Gamma_1$ and $\Gamma_2$ are determined by matrix elements at the contacts[30], which may vanish in the presence of destructive interference at the molecule-electrode interface.

**Example 3. Fano resonances.**

In this third example of QI, we examine how destructive quantum interference can be used to control molecular-scale transport. The control of electron transport by manipulating destructive interference was discussed in [31,32], where it was shown that Fano resonances can control electrical transport and lead to giant thermopowers and figures of merit in single-molecule devices.

Fano resonances occur when a bound state is coupled to a continuum of states. For a single molecule connected to metallic electrodes, the continuum of states is supplied by the metal and therefore, not surprisingly, this type of destructive interference is usually not present in the isolated molecule. In [32], it was shown that when a 'pendant' orbital of energy $\varepsilon_p$ is coupled to the above Breit-Wigner resonance via a coupling matrix element α, equation (13) is replaced by

$$T(E) = 4\Gamma_1 \Gamma_2/[(E - \varepsilon)^2 + (\Gamma_1 + \Gamma_2)^2] \qquad (14)$$

where $\varepsilon = \varepsilon_n + \alpha^2/(E - \varepsilon_p)$. This formula shows that when the electron anti-resonates with the pendant orbital (ie when $E = \varepsilon_p$), $\varepsilon$ diverges and electron transmission is destroyed. It also shows that a resonance occurs when $E - \varepsilon = 0$. Ie when $(E - \varepsilon_n)(E - \varepsilon_p) + \alpha^2 = 0$. There are two solutions to this quadratic equation. The solution close to $E = \varepsilon_n$ is the Breit-Wigner resonance, whereas the solution close to $E = \varepsilon_p$ is a resonance close to the anti-resonance. This means that a Fano resonance is composed of both an anti-resonance and a nearby resonance.

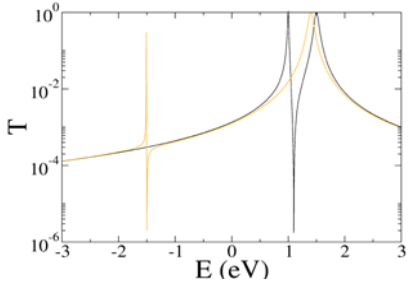

Figure 5. A plot of equation 2, for two different choices of $\varepsilon_1$. For the black curve $\varepsilon_p = 0.9$. For the brown curve $\varepsilon_p = -1.5$. The BW resonance occurs at $E = \varepsilon_n = 1.5$.

The vanishing of electron transmission due to a Fano resonance is particularly dramatic and can be distinguished from more general forms of destructive interference, such as multiple-path quantum interference, because the latter is sensitive to the changes in the positions of the electrodes, whereas the former is not. In the latter case, if for a specific location of electrode, $\Gamma_2 = 0$, then $\Gamma_2$ can be made non-zero by moving the electrode to a nearby location. On the other hand, when $E = \varepsilon_p$, equ. (2) shows that $T(E)$ vanishes no matter how large or small the values of $\Gamma_1$ or $\Gamma_2$ and provided equ. (14) remains valid, this cannot be remedied by adjusting the values of $\Gamma_1$ or $\Gamma_2$.

Finally, it should be noted that since a Fano resonance consists of a 'dip' at $E = \varepsilon_p$, and a 'peak' at $E = \varepsilon$, then if the Fano resonance appears near the Fermi energy and if its position fluctuates rapidly on the time scale of a measurement, possibly due to environmental fluctuations, then the Fano 'peak' will dominate the average conductance, rather than the 'dip', because the average of a large and a small positive number is dominated by by the larger number. [For example $(1+10^{-6})/2$ is approximately ½.]

As an example of real-life Fano resonances, it is known[31-33,36,41] that pendant oxygens promote destructive quantum interference[47,48] due to the creation of a localized state in the central part of the molecule, whose effect is similar to that of a side group state[41,36], and create sharp anti-resonances near the Fermi energy $E_F$, leading to a large suppression in electrical conductance[37-40] and an increase in thermopower.

**Example 4. A multi-branched 'molecule'.**

As an example of a more complex structure exhibiting QI, which can be solved analytically and contains the above three forms of QI as special cases, consider the multi-branched structure shown in figure 6, which is composed of (generally different) left and right leads connected to a structure containing M (generally different) branches.

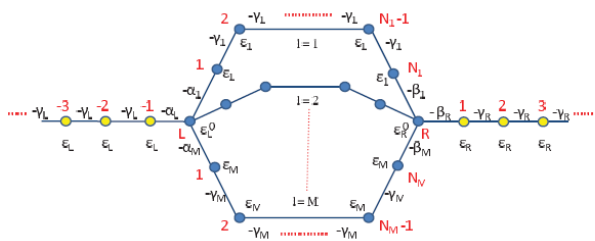

Figure 6. A multi-branch structure with nodal sites L and R (on the left and right) connecting external current-carrying leads, by hopping matrix elements $-\alpha_L$ (on the left) and $-\beta_R$ (on the right), and to internal branches (l), by hopping

matrix elements $-\alpha_l$ and $-\beta_l$, respectively. The energies of the nodal sites are $\varepsilon^0{}_L$ and $\varepsilon^0{}_R$. The site energy and hopping matrix element of branch $l$ are $\varepsilon_l$ and $-\gamma_l$, respectively.

An analytic formula for the transmission coefficient of the above structure is presented in [29], where it is shown that the transmission coefficient is given by

$$T(E) = v_L \left(\frac{\alpha_L}{\gamma_L}\right)^2 |G_{RL}|^2 \left(\frac{\beta_R}{\gamma_R}\right)^2 v_R \qquad (15)$$

In this expression, $v_L$ ($v_R$) is the electron group velocity in the left (right) lead, $\gamma_L$ ($\gamma_R$) is the hopping element in the left (right) lead, $\alpha_L$ ($\beta_R$) are the coupling between the left (L) and right (R) nodal atom to the left (right) lead and $G_{RL}$ is the Greens function of the whole structure describing a wave propagating from nodal atom L to nodal atom R. Clearly, this formula captures the intuitively-obvious property that for a wave to propagate from the left to the right lead, the electron velocities in the leads, their coupling to the nodal atoms and $G_{RL}$ must all be non-zero.

Appendix 1 (below) provides a simple MATLAB code, which evaluates this formula and provides plots of the transmission coefficient versus energy for user-selected, many-branched structures. It also evaluates the individual currents in each of the branches. Interestingly, although these currents must add up to the total current through the device, the individual currents in some branches can be greater than the total current, because the currents in other branches can be in the opposite direction. By modifying this few lines of MATLAB, the reader can have a little fun and learn about quantum interference by evaluating the transmission coefficient of a variety of different model systems.

To utilise equation (15), the hopping elements $\gamma_L, \gamma_R, \gamma_l$ and orbital energies $\varepsilon_L, \varepsilon_R, \varepsilon_l$ defining the left (L) and right (R) leads and each branch $l$ should be chosen. For a given energy $E$, the wavevectors in L, R and $l$ are then given by $k_L(E) = \cos^{-1}(\varepsilon_L - E)/2\gamma_L$, $k_R(E) = \cos^{-1}(\varepsilon_R - E)/2\gamma_R$ and $k_l(E) = \cos^{-1}(\varepsilon_l - E)/2\gamma_l$. The sign of the wave vectors is chosen such that the corresponding group velocities $v_L = 2\gamma_L \sin k_L(E)$, $v_R = 2\gamma_R \sin k_R(E)$ and $v_l = 2\gamma_l \sin k_l(E)$ are positive, or if the wavevector is complex, such that the imaginary part is positive. Next the orbital energies $\varepsilon^0{}_L, \varepsilon^0{}_R$ of the nodal sites L and R and their respective couplings $-\alpha_L, -\alpha_l$ and
$-\beta_R, -\beta_l$ to the leads and branches should be chosen.

The final step in evaluating equation (15) is to compute the Greens function $G_{RL}$ connecting the left nodal site $L$ to the right nodal site $R$ via the expression:

$$G_{RL} = y/\Delta \qquad (16)$$

In this equation, the numerator $y$ contains information about the branches and their couplings to the nodal sites only (ie it contains no information about the left and right leads) and describes one contribution to quantum interference due to the multiple paths within the structure of figure 6. It is given by the following superposition of contributions from each of the $M$ branches:

$$y = \sum_{l=1}^{M} y_l \qquad (17)$$

where $y_l$ describes the ability of a wave to propagate from the left to the right nodal site and is given by

$$y_l = \frac{\alpha_l \beta_l \sin k_l}{\gamma_l \sin k_l(N_l + 1)} \qquad (18)$$

where $N_l$ is the number of atoms in branch $l$. (For the special case $N_l = 1$, one should choose $\alpha_l = \beta_l = \gamma_l$.)

This expression immediately reveals an interesting rule for combining conductors in parallel, which is valid in the limit that $\Delta$ does not depend on the number or nature of the branches. (This approximation will be discussed further below.) In this case, for a structure containing a single branch $l$, equation (15) yields for the transmission coefficient $T_l(E)$

$$T_l(E) = B(y_l)^2$$

where $B = v_L (\frac{\alpha_L}{\gamma_L})^2 |\Delta|^{-2} (\frac{\beta_R}{\gamma_R})^2 v_R$

On the other hand in the presence of two branches ($l=1$ and $l=2$), the transmission coefficient is

$$T(E) = B(y_1 + y_2)^2 = T_1(E) + T_2(E) + 2A y_1 y_2$$

If $y_1$ and $y_2$ are of the same sign, this yields $(E) = T_1(E) + T_2(E) + 2\sqrt{T_1(E) T_2(E)}$. From equation (4) since the corresponding conductances (G, $G_1$ and $G_2$) are obtained by multiplying the transmission coefficients by the quantum of conductances, this means that the conductances satisfy

$$G = G_1 + G_2 + 2\sqrt{G_1 G_2}. \qquad (19)$$

If the branches are identical, so that $G_1 = G_2$, then one finds $G = 4G_1$, instead of the classical result obtained from Kirchoff's law for two conductors in parallel of $G = 2G_1$. This expression was derived in [42]. For three branches in parallel, the corresponding expression is clearly

$$G = (\sqrt{G_1} + \sqrt{G_2} + \sqrt{G_3})^2 \qquad (20)$$

However, it should be emphasised that this expression is not general, because the condition of "constant $\Delta$" is not universally satisfied and in general the quantities $y_l$ do not have the same sign.

To understand more precisely how parallel branches combine to yield the overall transmission coefficient, we need to evaluate the denominator $\Delta$ of equation (16), which is given by

$$\Delta = y^2 - (a_L - x_L)(a_R - x_R) \qquad (21)$$

In this expression, the quantities $x^L$ and $x^R$ describe how a wave from the left or right nodal sites is reflected back to those sites and are given by

$$x^L = \sum_{l=1}^{M} x_l^L \qquad (22)$$

$$x^R = \sum_{l=1}^{M} x_l^R \qquad (23)$$

where

$$x_l^L = \frac{\alpha_l^2 \sin k_l(N_l)}{\gamma_l \sin k_l(N_l+1)} \qquad (24)$$

and $\qquad x_l^R = \frac{\beta_l^2 \sin k_l(N_l)}{\gamma_l \sin k_l(N_l+1)} \qquad (25)$

Finally, the quantities $a_L$ and $a_R$ contain information about the nodal site energies and their coupling to the left and right leads and are given by

$$a_L = (\varepsilon_L^0 - E) - \frac{\alpha_L^2}{\gamma_L} e^{ik_L} \qquad (26)$$

and

$$a_R = (\varepsilon_R^0 - E) - \frac{\beta_R^2}{\gamma_R} e^{ik_R} \qquad (22)$$

Equation (21), reveals that the condition of "constant $\Delta$" is approximately satisfied if $y$, $x_L$ and $x_R$ are small compared with $a_L$ and $a_R$, which means that the branches should be coupled only weakly to the nodal sites.

Equation (15) contains a great deal of information about manifestations of quantum interference in a variety of contexts. For example in [29], it is shown how the Breit-Wigner formula (13) is obtained from this equation as special case and how odd-even effects arise in linear chains of atoms. As another example, we now derive equation (11) for the Greens function $G_{RL}$ of the rings of atoms shown in figure 4b and 4c. In this case, there are two identical branches with $N_1$ atoms in branch 1 and $N_2$ atoms in branch 2. For a para-connected phenyl ring, $N_1 = N_2 = 2$, while for a meta connect ring, $N_1 = 1$ and $N_2 = 3$. Since all atoms are identical, we approximate the system by choosing all site energies to be equal to a constant $\varepsilon_0$ and all couplings (except $\alpha_R$ and $\alpha_L$) equal to $\gamma$, ie $\alpha_l = \beta_l = \gamma_l = \gamma$. This means that all wave vectors are equal to $k(E) = \cos^{-1}(\varepsilon_0 - E)/2\gamma$ and $x_L = x_R$.

First consider the case of an isolated ring (figure 4a) for which $\alpha_L = \beta_R = 0$, in which case $a_L = a_R = 2\gamma \cos k$, $x_l = \frac{\gamma \sin k_l(N_l)}{\sin k_l(N_l+1)}$, $y_l = \frac{\gamma \sin k_l}{\sin k_l(N_l+1)}$. Since $\gamma \cos k - x_l = \gamma \sin k \frac{C_l}{S_l}$ where $S_l = \sin k_l(N_l+1)$ and $C_l = \cos k_l(N_l+1)$, one obtains $a_L - x = \gamma \sin k \left(\frac{C_1}{S_1} + \frac{C_2}{S_2}\right)$, $y = \gamma \sin k (S_1 + S_2)/S_1 S_2$ and $\Delta = \frac{4\gamma^2 \sin^2 k}{S_1 S_2} \sin^2 kN/2$, where $N = N_1 + N_2 + 2$. These combine to yield

$$G_{RL} = \frac{y}{\Delta} = \frac{\cos k\left(\frac{N_1 - N_2}{2}\right)}{2\gamma \sin k \sin kN/2} \qquad (26)$$

Which is identical to equation (8), because with the notation of figure 4a, for any choice of $i$ and $j$ in equation (8) $|i-j|=N_1+1$ and $N=N_1+N_2+2$. More generally, when the coupling to the left and right leads ($\alpha_L$ and $\beta_R$) are not zero, $a_L = 2\gamma \cos k + \sigma_L$ where $\sigma_L = (\varepsilon_L^0 - \varepsilon_0) - \frac{\alpha_L^2}{\gamma_L} e^{ik_L}$ and similarly for $a_R$. In this case, we obtain equation (11)

$$G_{RL} = \frac{\cos k\left(\frac{N_1 - N_2}{2}\right)}{2\gamma \sin k \sin \frac{kN}{2} + \sigma_{ring}} \qquad (27)$$

where $\sigma_{ring} = \frac{2\gamma \sin k \sin kN (\sigma_L + \sigma_R) - S_1 S_2 \sigma_L \sigma_R}{2\gamma \sin k \sin \frac{kN}{2}} \qquad (28)$

Furthermore, the calculation can easily be repeated for a single branch to yield

$$G_{RL} = \frac{-\sin k}{\gamma \sin k(N_1+3) + \sigma_{chain}} \qquad (29)$$

where

$$\sigma_{chain} = -2 \sin k(N_1 + 2)(\sigma_L + \sigma_R) - \sin k(N_1 + 1) \sigma_L \sigma_R / \gamma \quad (30)$$

(Note that in the notation of equation (12), $N = N_1 + 2$).

As an example, for $N=6$, $k=\pi/2$, equation (28) for the a ring yields

$$G_{RL} = \frac{-2\gamma \cos k(\frac{N_1-N_2}{2})}{4\gamma^2 - \sin k(N_1+1)\sin k(N_2+1)\sigma_L\sigma_R} \quad (30)$$

For the para case, where $N_1=N_2=2$, this yields

$$G_{RL} = \frac{-2\gamma}{4\gamma^2 - \sigma_L\sigma_R} \quad (31)$$

For the meta case, where $N_1=1$, $N_2=3$, it yields $G_{RL} = 0$ and for the ortho case, where $N_1=0$, $N_2=4$, it yields

$$G_{RL} = \frac{2\gamma}{4\gamma^2 - \sigma_L\sigma_R} \quad (32)$$

These expressions demonstrate that at the centre of the HOMO-LUMO gap, ortho and para couplings lead to the same electrical conductance.

As a second example, of this odd-even conductance variation as a function of $N_1$, consider the Greens function of a linear chain at k=π/2. In this case equation (29) yields

$$G_{RL} = (-1)^{\frac{N_1+1}{2}} \frac{1}{2(\sigma_L+\sigma_R)} \quad \text{for } N_1 \text{ odd}$$

and

$$G_{RL} = (-1)^{\frac{N_1}{2}} \frac{1}{\gamma+\sigma_L\sigma_R/\gamma} \quad \text{for } N_1 \text{ even}$$

which shows that the conductance of such a chain also exhibits an odd-even oscillation as a function of the chain length. This shows that well-known conductance oscillations in atomic chains[43] are closely related to meta versus para conductance variations found in aromatic rings.

The formula (15) is rather versatile, because the precise meaning of the symbols depends on context. For examples the 'sites' within the scatterer could refer to molecular orbitals or orbitals localised on groups of atoms within a molecule. As an example of this versatility, consider the case of a single branch (M=1) containing zero sites ($N_1=0$) and choose $\alpha_1 = \beta_1 = \gamma_1$, $\beta_R = \gamma_R$, $\varepsilon^0_R = \varepsilon_R$, $\varepsilon^0_L = E_n$. In this case, the nodal site L plays the role of a molecular orbital weakly coupled to the leads and figure 6 reduces to figure 7a below, for which y= $\alpha_1$, x=0, $a_R = \varepsilon_R - E - \gamma_R \exp(ik_R) = 2\gamma_R \cos(k_R) - \gamma_R \exp(ik_R) = \gamma_R \exp(-ik_R)$ and $a_{LR} = E_n - E - \frac{\alpha_L^2}{\gamma_L}\exp(ik_L)$. Substitution into equations (15) and (16) yields the Breit-Wigner formula (13), with $\Sigma = \frac{\alpha_L^2}{\gamma_L}\cos(k_L) + \frac{\alpha_1^2}{\gamma_L}\cos(k_R)$, $\Gamma_1 = \frac{v_L\alpha_L^2}{2\gamma_L^2} = \frac{\alpha_L^2}{\gamma_L}\sin(k_L)$ and $\Gamma_2 = \frac{\alpha_1^2}{\gamma_R}\sin(k_R)$. As a final example, the choice M=2, $N_1=0$, $\alpha_1 = \beta_1 = \gamma_1$, $\beta_R = \gamma_R$, $\varepsilon^0_R = \varepsilon_R$, $\varepsilon^0_L = E_n$, $N_2=1$, $\beta_2 = 0$, $\alpha_2 = \alpha$ yields the structure shown in figure 7b, in which a pendant orbital of energy $\varepsilon_p$ is weakly coupled to the orbital $E_n$. In this case, equations (15) and (16) combine to yield equation (14) describing a Fano resonance.

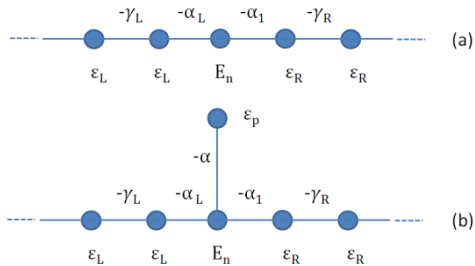

Figure 7. (a) A special case of figure 6, obtained by setting M=1, $N_1=0$, $\alpha_1 = \beta_1 = \gamma_1$, $\beta_R = \gamma_R$, $\varepsilon^0_R = \varepsilon_R$, $\varepsilon^0_L = E_n$. (b) A special case of figure 6, obtained by setting M=2, $N_1=0$, $\alpha_1 = \beta_1 = \gamma_1$, $\beta_R = \gamma_R$, $\varepsilon^0_R = \varepsilon_R$, $\varepsilon^0_L = E_n$, $N_2=1$, $\beta_2 = 0$, $\alpha_2 = \alpha$.

**D. Thermoelectricity and the role of inter-scatterer quantum phase averaging.**

The transmission coefficient controls not only the electrical conductance of single molecules, but also a range of other transport properties. In this section, we discuss issues associated with thermoelectricity. The thermopower or Seebeck coefficient (*S*) and thermoelectric figure of merit (*ZT*) of a material or of a nanojunction are defined as $S = -\Delta V/\Delta T$ and $ZT = S^2 GT/\kappa$ where $\Delta V$ is the voltage difference created between the two ends of the junction when a temperature difference $\Delta T$ is established between them, *G* is the electrical conductance, *T* is the ambient temperature and $\kappa$ is the thermal conductance. Despite several decades of development, the best inorganic thermoelectric materials, such as bismuth telluride (Bi2Te3)-based alloys, possess a figure of merit *ZT* close to unity only, which is not sufficient to create a viable technology platform for harvesting waste heat. As an alternative, organic thermoelectric materials are now being investigated and are already showing promising values of both *ZT* and thermopower.

A key strategy for improving the thermoelectric properties of inorganic materials has been to take advantage of nanostructuring[44-46], which leads to quantum confinement of electrons and enhanced thermoelectric performance[46-52]. The single-molecule building blocks of organic materials offer the ultimate limit of electronic confinement, with quantised energy level spacings, which are orders of magnitude greater than room temperature. Therefore it is natural to examine the thermoelectric performance of single-molecule junctions as a stepping stone towards the design of new materials. The ability to measure thermopower in single-molecule junctions is relatively new[53-58] and the thermoelectric properties of only a few molecules have been measured.

To calculate thermoelectric quantities[59] it is useful to introduce the non-normalised probability distribution *P(E)* defined by

$$P(E) = -T(E)\frac{\partial f(E)}{\partial E},$$

where *f(E)* is the Fermi-Dirac function, whose moments are denoted

$$L_i = \int_{-\infty}^{\infty} dE\, P(E)(E - E_F)^i, \qquad (33)$$

where $E_F$ is the Fermi energy. With this notation, the low-bais conductance, *G* is

$$G(T) = \frac{2e^2}{h} L_0, \qquad (34)$$

where *e* is the electronic charge, *h* is the Planck constant, and *T* is the temperature. The thermopower *S* is

$$S(T) = -\frac{1}{eT}\frac{L_1}{L_0}, \qquad (35)$$

the electronic contribution to the thermal conductance $\kappa_{el}$ is

$$\kappa_{\text{el}} = \frac{2}{h}\frac{1}{T}\left(L_2 - \frac{L_1^2}{L_0}\right), \qquad (36)$$

and the electronic contribution to the figure of merit $Z_{el}T$ is

$$Z_{\text{el}}T = \frac{1}{\frac{L_0 L_2}{L_1^2} - 1}. \qquad (37)$$

As an aside, for $E$ close to $E_F$, if $T(E)$ varies only slowly with $E$ on the scale of $k_B T$ then these expressions take the well-known forms:

$$G(T) \approx \left(\frac{2e^2}{h}\right) T(E_F), \qquad (38)$$

$$S(T) \approx -\alpha e T \left(\frac{d\, lnT(E)}{dE}\right)_{E=E_F}, \qquad (39)$$

$\kappa \approx L_0 T G$, where $\alpha = \left(\frac{k_B}{e}\right)^2 \pi^2/3$ is the Lorentz number.

The above approximate formulae suggest a number of strategies for enhancing the thermoelectric properties of single molecules. For example equation (9) demonstrates the "rule of thumb" that S is enhanced by increasing the slope of ln $T(E)$ near $E=E_F$ and therefore the presence of sharp resonances near $E=E_F$ are of interest. For the purpose of increasing the sharpness of such resonances, Fano resonances appear to be more attractive than Breit-Wigner resonances, because the width of the latter is governed by the parameters $\Gamma_1$ and $\Gamma_2$, whereas the width of the former is determined by the parameter α in equation (14). To achieve a stable junction, it is desirable that there is strong coupling to the electrodes, so the parameters $\Gamma_1$ and $\Gamma_2$ should as large as possible and therefore in practice, Breit-Wigner resonances are likely to be rather broad. On the other hand the parameter $α$ is an intra-molecular coupling, which is independent of the coupling to the electrodes and can be tuned without affecting junction stability.

The above formulae provide guidelines for designing new thermoelectric materials from the single molecule upwards. To illustrate such a strategy, consider two quantum scatterers (labelled 1 and 2) in series, whose transmission and reflection coefficients are $T_1$, $T_2$ and $R_1$, $R_2$ respectively. It can be shown that the total transmission coefficient for the scatterers in series is $T = \frac{T_1 T_2}{1 - 2\sqrt{R_1 R_2} \cos\varphi + R_1 R_2}$, where φ is a quantum phase due to QI between the scatterers[60,61]. Experimentally-quoted conductances are identified usually with the most probable value of $\log_{10}\frac{G}{G_0}$ and if this possesses a Gaussian distribution, then it equates to the ensemble average of $\log_{10}\frac{G}{G_0}$, which we denote by $\overline{\log_{10}T}$. From the above expression for T,

$\overline{\log_{10}T} = \overline{\log_{10}T_1} + \overline{\log_{10}T_2} + \overline{\log_{10}(1 - 2\sqrt{R_1 R_2}\cos\varphi + R_1 R_2)}$. If the phase $\varphi$ is uniformly distributed between 0 and $2\pi$, then the third term on the right hand side averages to zero, because of the mathematical identity $\int_0^{2\pi} d\varphi \, \overline{\log_{10}(1 - 2\sqrt{R_1 R_2}\cos\varphi + R_1 R_2)} = 0$. Hence all information about the inter-scatterer quantum phase is lost and $\overline{\log_{10}T} = \overline{\log_{10}T_1} + \overline{\log_{10}T_2}$, or equivalently $\frac{G_{Total}}{G_0} = \frac{G_1}{G_0}\frac{G_2}{G_0}$. More importantly, if $S_1$ and $S_2$ are the thermopowers of 1 and 2 respectively, then according to equ (39) the combined thermopower of the new molecule is $S = S_1 + S_2$.

This strategy for designing new materials was demonstrated recently in [62], where it was shown that the thermopower of two $C_{60}$s coupled in series is approximately twice that of a single $C_{60}$.

## Conclusions

Rather than provide a review if the literature, the aim of this tutorial has been to gather together some basic concepts and mathematical tools, which underpin the fundamentals of phase-coherent electron transport through single molecules. The key quantity of interest has been the transmission coefficient $T(E)$, which yields the electrical conductance and various thermoelectric coefficients. Since $T(E)$ is strongly affected by QI, it provides us with strategies for utilising QI to increase the performance of molecular-scale devices. The MATLAB code in the appendix provides a route to exploring QI in multi-branched structures described by a tight-binding Hamiltonian.

The GOLLUM, TRANSIESTA and TURBOMOLE codes provide a method of exploring such effects at the level of density functional theory. These codes can be used to explore different realisations of the concepts discussed in this tutorial. Single-molecule electronics is a field very much in its infancy. The optimal combinations of electrodes, anchor groups and functional units for delivering enhanced performance are still largely unknown. The concepts discussed in this tutorial, when combined with synthesis of new molecules and experimental exploration of single-molecule devices structures should help to underpin future strategies for addressing some of these challenges.

## Learning points:

1. The central role of the transmission coefficient T(E) in determining electrical conductances and thermoelectric coefficients.
2. Three manifestations of QI in single-molecules, namely Mach-Zehnder interferometry, Breit-Wigner resonances and Fano resonances.
3. Strengths and weaknesses of density functional theory.
4. A simple tight-binding model of para-, meta- and ortho-linked rings.
5. An ability to model multi-branched structures.

### Acknowledgements
This article is based on many fruitful interactions with experimental colleagues in the FUNMOLS and MOLESCO ITNs and theoretical colleagues in a 20-years series of EU networks, of which the latest is the NanoCTM ITN. In particular I would like to thank Thomas Wandlowskii, Martin Bryce, Richard Nichols, Simon Higgins, Nicholas Agrait, Harry Anderson and Andrew Briggs for many fruitful discussions about experimental design, measurement and synthesis and Jaime Ferrer for theoretical discussions and co-development of GOLLUM.  This work is supported by the UK EPSRC and the EU ITN MOLESCO.

**Appendix: A simple MATLAB program to evaluate equ.(5) and equ (19) of Sparks, Garcia-Suarez, Manrique and Lambert (SGML)** [19]

Below is a simple MATLAB formula for evaluating the SGML formula [19] and also equation (19) of ref [19] for the currents within the individual branches of the multi-branch structure shown in figure 6.

Enjoy exploring QI in a variety of systems by varying the input parameters!

```
%%%Start of program
%%%%% Choose the parameters defining the  system
gammaL = 1.1; % gamma of left lead
epsiL=0;    % epsi of left lead
alphaL = 0.5; %coupling to left lead
epsi0L=0;   %site energy of left node
gammaR = 1.1; %gamma of right lead
epsiR=0;   %epsi of right lead
betaR = 0.5; %coupling to right lead
epsi0R=0;  %site energy of right node
m=2;       %number of branches
alpha=ones(1,m); %array of left coupling to branches
beta=ones(1,m);  %array of right coupling to branches
gamma=ones(1,m); %array of gammas of branches
epsi=zeros(1,m);  %array of site energies of branches
n=2*ones(1,m); %array containing the number of atoms in each branch

% In the above array 'n' both branches have 2 atoms so this is a
% para-connected ring with 6 atoms
% For a meta-connected ring with 6 atoms, add the following two %lines (ie remove the %s)
% n(1,1)=1
% n(1,2)=3

%%%%% Choose a set of energies for graph plotting
nenergies=10000;
energies=ones(1,nenergies);
trans=ones(1,nenergies);
current=ones(m,nenergies);
```

```matlab
emin=-2*max(gammaL,gammaR);
emax=2*max(gammaL,gammaR);
deltae=(emax-emin)/nenergies;

%%%% Loop over all energies
 for count=1: nenergies;
    E=emin+deltae*count;
    energies(1,count)=E;

kL=acos((epsiL-E)/(2*gammaL)); % wavevector of left lead
if abs((epsiL-E)/(2*gammaL)) <=1
    vL=2*gammaL*sin(kL);       % group velocity of left lead
else
    vL=0;
    disp('no open channel in left lead')
end
kR=acos((epsiR-E)/(2*gammaR)); % wavevector of left lead
if abs((epsiR-E)/(2*gammaR)) <= 1
    vR=2*gammaR*sin(kR);       % group velocity of left lead
else
    vR=0;
    disp('no open channel in right lead')
end
k=acos((epsi-E)./(2*gamma));
v=2*gamma.*sin(k);
yl=alpha.*beta.*sin(k)./(gamma.*sin(k.*(n+1)));
y=sum(yl);
xLl=alpha.*alpha.*sin(k.*n)./(gamma.*sin(k.*(n+1)));
xL=sum(xLl);
xRl=beta.*beta.*sin(k.*n)./(gamma.*sin(k.*(n+1)));
xR=sum(xRl);
aL=(epsi0L-E)-(alphaL^2/gammaL)*exp(i*kL);
aR=(epsi0R-E)-(betaR^2/gammaR)*exp(i*kR);
del=y*y-(aL-xL)*(aR-xR);
GRL=y/del;
transmission=vL*vR*(alphaL*betaR/(gammaL*gammaR))^2*abs(GRL)^2;
trans(1,count)=transmission;
current(:,count)=transmission*yl'./y;
end

 %%%% Plot the graphs
subplot(2,2,1)
plot(energies,trans)
title(['T    (','m = ',num2str(m),')'])
subplot(2,2,2)
plot(energies,current(1,:),'k')
title(['current in lead 1  ','(',' n1 = ',num2str(n(1,1)),')'])
if m >= 2
    subplot(2,2,3)
plot(energies,current(2,:),'r')
title(['current in lead 2  ','(',' n2 = ',num2str(n(1,2)),')'])
end
if m >= 3
    subplot(2,2,4)
plot(energies,current(3,:),'g')
title(['current in lead 3  ','(',' n3 = ',num2str(n(1,3)),')'])
end
```

## Notes and references